\documentclass[aps,preprint,pre,superscriptaddress,amsmath,runinaddress,showpacs]{revtex4}
\usepackage{graphicx}
\usepackage{subfigure}
\usepackage{psfrag}
\usepackage{bm}

\begin{document}

\title{Dynamics of strongly correlated ions in a partially ionized quantum plasma}

\author{M.~Bonitz}
\author{P.~Ludwig}
\affiliation{Institut f\"{u}r Theoretische Physik und Astrophysik, Christian-Albrechts Universit\"{a}t zu Kiel, 24098 Kiel, Germany}
\author{J.W.~Dufty}
\affiliation{Physics Department, University of Florida, Gainesville, FL}

\date{\today}

\begin{abstract}
A scheme which allows to compute the dynamics of strongly correlated classical ions embedded into a partiallzy ionized quantum plasma by first principles molecular dynamics is presented. The dynamically screened dust approach of Joyce and Lampe [Phys. Rev. Lett. {\bf 88}, 095006 (2002] ist generalized to quantum systems. The electrons are treated fully quantum-mechanically taking into account their dynamical screening of the ion-ion interaction in linear response on the basis of an extended Mermin formula. The scheme allows to include the effect of the electron dynamics, electron streaming, wake effects and electron magnetization.
\end{abstract}


\maketitle

\section{Introduction}
Strong correlation effects in ensembles of charged particles are of high importance in many fields of physics, including plasmas, the electron gas in solids or electron-hole plamas, e.g. \cite{dubin99,bonitz2008pop} and references therein. In recent years charged particles spatially confined in trapping potentials have attracted considerable interest. Examples are ultracold ions \cite{Wineland1987,drewsen1998}, dusty plasmas \cite{Arp2004,bonitz2006}, or electrons in quantum dots \cite{Alex2001}, 
see Ref.~\cite{bonitz2008pop} for an overview.

\section{Basic equations}\label{sec:eqsl}
The system hamiltonian is given by
\begin{eqnarray}
{\hat H} &=& \sum_a {\hat H}_a + \sum_{ab} {\hat H}_{ab}, \\
{\hat H}_a &=& {\hat H}^0_a + {\hat H}^{int}_a,\\
{\hat H}^0_a &=& \sum_{i=1}^{N_a}\frac{1}{2m_{a}}\left( \frac{\hbar}{i}\nabla_i - \frac{e_a}{c}{\bf A}({\bf r}_i,t) \right)^2 + e_a\phi({\bf r}_i,t), \\
{\hat H}^{int}_a &=& \sum_{1\le i < j\le N_a} V_{aa}(|{\bf r}_i-{\bf r}_j|), \\
{\hat H}_{ab} &=& \sum_{1\le i \le N_a}\sum_{1\le j \le N_b} V_{ab}(|{\bf r}_i-{\bf r}_j|),
\label{h}
\end{eqnarray}
where ${\bf A}$ and $\phi$ denote the vector and scalar potential of an external electromagnetic field.
We will consider situations where the heavy particles, i.e. ions and neutrals are classical. Extensions to weak ion degeneracy will be briefly discusseed below in Sec.~\ref{q-ion-pot_ss}.
Also, the neutrals will be assumed to have a minor effect on the ions which can be neglected at this stage. The effect of the neutrals on the electron dynamics will be included by an effective collision frequency, $\nu_{en}=\sum_k \nu^{k}_{en}$, summing over all species of neutrals.

The nonequilibrium dynamics of electrons (e) and ions (i) can be described by reduced density operators and is given by \cite{bonitz-book}
\begin{eqnarray}
 i\hbar \frac{\partial {\hat F}_a(1)}{\partial t} - [{\hat H}_a(1), {\hat F}_a(1)] = 
\sum_b n_b Tr_2[{\hat V}_{ab}(1,2), {\hat F}_{ab}(1,2)], 
\label{f1_eq}
\end{eqnarray}
where $a=e,i$, and ${\hat F}_{ab}(1,2)$ is the binary density operator and the operators are normalized to the volume according to \cite{bonitz-book}
\begin{equation}
{\rm Tr}_{1\dots s} F_{1\dots s} = {\cal V}^s.
\end{equation}. 
In the following, quantum exchange effects will be irrelevant and are neglected. Then we may introduce the binary correlation operator ${\hat g}_{ab}(1,2)$ by separating the ideal contribution, ${\hat F}_{ab}(1,2)={\hat F}_{a}(1){\hat F}_{b}(2)+{\hat g}_{ab}(1,2)$.

While the ions are classical, no restrictions on the strength of their coupling are being imposed. In fact, we are particularly interested in strong ion-ion correlations leading to fluid and solid-like many-particle behavior. Therefore, the ion-ion interaction will be treated exactly. 
On the other hand, electron-ion coupling can be assumed weak, due to the quantum degeneracy (delocalization) of the electrons. Therefore, it will be sufficient to treat e-i interactions on the mean-field level, so e-i correlations will be neglected, 
${\hat g}_{ei}\approx 0$. Then the ion equation becomes
\begin{eqnarray}
 i\hbar \frac{\partial {\hat F}_i(1)}{\partial t} - [{\hat H}_i(1), {\hat F}_i(1)] = 
n_i Tr_2[{\hat V}_{ii}(1,2), {\hat F}_{ii}(1,2)] + n_e Tr_2[{\hat V}_{ie}(1,2), {\hat F}_{e}(2){\hat F}_{i}(1)]+ 
\nonumber\\
n_n Tr_2[{\hat V}_{in}(1,2), {\hat F}_{in}(1,2)].
\label{fi_eq}
\end{eqnarray}
The second term on the right gives rise to an effective mean field (Hartree potential) for the ions which is created by the electrons, 
${\hat W}_{ie}(1)=n_e Tr_2 {\hat V}_{ie}(1,2) {\hat F}_{e}(2)$. The last term describes the interaction of ions with neutrals.

Since the ions are classical we now take the semiclassical limit in the equation (\ref{fi_eq}), first transforming to the Wigner representation \cite{bonitz-book}. Then the ion density operator is replaced by the Klimontovich phase space density \cite{klimontovich75} 
\begin{equation}
{\hat F}_i \rightarrow N_i({\bf r},{\bf p},t)=\sum_{k=1}^{N}\delta[{\bf r}-{\bf r}_k(t)]\delta[{\bf p}-{\bf p}_k(t)] + O(\hbar^2),  
\label{n_klim}
\end{equation}
which still includes all fluctuations. This function obeys a Liouville type equation
\begin{eqnarray}
 \left\{ \frac{\partial}{\partial t} + \frac{{\bf p}}{m}\frac{\partial}{\partial {\bf r} } - 
\frac{\partial}{\partial {\bf r} }\left(V^Q_i+W_{ie}+W_{in}+e_i\phi^{ext}\right)\frac{\partial}{\partial {\bf p}} \right\} 
N_i({\bf r},{\bf p},t) &=& 0.
\label{fi_eqcl}
\end{eqnarray}
Here $\phi^{ext}$ is the electrostatic potential due to all external fields and $W_{in}$ and $W_{ie}$ are the potential energies induced by the neutrals and electrons, respectively. $W_{ie}$ is obtained after an ensemble average of ${\hat W}_{ie}$ over the electron 
density operator, i.e. $W_{ie}=\langle {\hat W}_{ie} \rangle|_{F_e}=e_i\phi_{e}$. The potential $\phi_{e}$ is the solution of 
Poisson's equation and determined by the electron density via
\begin{equation}
 \phi_{e}({\bf r},t)=e\int d^3r' \frac{n_e({\bf r}',t)}{|{\bf r}-{\bf r}'|}.
\label{phi_e}
\end{equation}

Furthermore, $V^Q_{i}$ denotes the total potential energy due to all pair interactions of the ions. 
The superscript ``Q'' indicates that ionic quantum effects (which appear only at very short pair distances) have been taken into account by replacing the bare pair  potential by an effective quantum pair potential, $V^Q_{ii}$. In the case of Coulomb interaction $V^Q_{ii}$ is the improved Kelbg potential, see \cite{afilinov_pre04} and references therein. Here we will, instead, have a pair interaction which is dynamically screened by the electrons giving rise to a different quantum potential, see 
Sec.~\ref{q-ion-pot_ss}. 
A similar procedure may be applied to the external potentials in the case that they vary on small length scales comparable to the size (or thermal DeBroglie wave length) of an ion, e.g. \cite{fromm08}, but we will not write this explicitly.

Eq. (\ref{fi_eqcl}) is equivalent to a set of Newton's equations for the ions (which are its characteristics) which is directly verified by introducing the definition (\ref{n_klim}) into (\ref{fi_eqcl})
\begin{equation}
\frac{d{\bf p}_k}{dt}=-\frac{d}{d {\bf r}} (V^Q_i+W_{ie}+e_i\phi^{ext})\big|_{{\bf r}={\bf r}_k} - \nu_{in}{\bf p}_k +{\bf y}(t),
\quad k=1,\dots, N_i,
 \label{newton_i}
\end{equation}
which treat the ion interactions and the external fields exactly. In case of a time-dependent electromagnetic field, 
the r.h.s. will contain the ionic Lorentz force instead of the gradient of $\phi^{ext}$. The last two terms on the r.h.s. are related to the potential $W_{in}$ in Eq.~(\ref{fi_eqcl}) and allow to simulate the effect of the neutrals on the ions (friction term and Langevin noise term), but may be neglected in most cases, except for a verly low degree of ionization.
In addition, the effect of the electrons is 
taken into account via the induced potential $\phi_{e}({\bf r},t)$, Eq. (\ref{phi_e}). The quality of this potential depends on the level of treatment of the electronic density which is computed from the electron density operator.  We, therefore, now turn to the computation of ${\hat F}_e$.

\section{Quantum electron distribution function in linear response}\label{fe_s}

We now return to the equation of motion (\ref{f1_eq}) for the electron density operator. Since we consider a dense degenerate electron system with small value of the electron Brueckner parameter, $r_{se}={\bar r}_e/a_B < 1$, electron-electron correlations are small. The main contribution to the interaction between electrons will, therefore, arise from the electronic mean field, $W_{ee}$, which will be treated exactly. The additional correlation effects between electrons and from scattering with neutral particles (contained in the electron binary correlation operator) will be treated approximately, within a relaxation time approximation. This term is denoted as collision integral $I_e$, its form will be specified below.
\begin{eqnarray}
 i\hbar \frac{\partial {\hat F}_e(1)}{\partial t} - [{\bar H}_e(1), {\hat F}_e(1)] &=& I_e(1), \\
{\bar H}_e(1) &=& {\hat H}_e^0(1) + {\hat W}_{ee}(1), \\ 
{\hat W}_{ee}(1) &=& n_e Tr_2 V_{ee}(1,2) {\hat F}_e(2).
\label{fe_eq}
\end{eqnarray}

\subsection{Momentum representation for a spatially inhomogeneous system}
In the following we will consider the response of the electron system to a spatially inhomogenous electrostatic field. To do this, it is convenient to transfrom the abstract operator equation (\ref{fe_eq}) into the momentum representation using as basis function free electron states  $|k\rangle$. Thus, we split the single-particle hamiltonian into a field-free part, ${\hat h}^0$, and a field part, ${\hat U}$, ${\hat H}_e^0 ={\hat h}_e^0 +{\hat U}$, with
\begin{eqnarray}
{\hat h}_e^0 &=& -\frac{\hbar^2}{2m_e}\nabla^2,\\
{\hat h}_e^0 |k\rangle &=& \epsilon_k|k\rangle, \quad \mbox{with} \;\epsilon_k=(\hbar k)^2/2m_e.
\end{eqnarray}
In this basis, the operators transform into matrices (we drop the time-dependence): $\langle k| {\hat F}_e |k'\rangle  = {\cal V}^{-1}f_{k,k'}$, $\langle k| {\hat h}^0_e |k'\rangle  = \epsilon_{k}\delta_{k,k'}$,  
$\langle k| {\hat U} |k'\rangle  = U_{k,k'}$. The normalization condition is $2\sum_k f_{kk}(t)=N_e(t)$ where the factor $2$ arises from the spin summation.
Multiplying Eq. (\ref{fe_eq}) from left by $\langle k|$ and right 
by $|k'\rangle$ it becomes a matrix equation
\begin{eqnarray}
 i\hbar \frac{\partial f_{k,k'}}{\partial t} - (\epsilon_k - \epsilon_{k'})f_{k,k'} 
-\langle k|[{\hat U}^{{\rm eff}},{\hat F}_e] |k'\rangle
&=& I_{k,k'}, 
\label{fkk_eq}
\end{eqnarray}
where the effective potential operator is ${\hat U}^{{\rm eff}}={\hat U}+{\hat W}_{ee}$. The evaluation of the commutator term will be performed below in linear response. For the full result, see Ref. \cite{bonitz-book}.

\subsection{Electron equation in linear response}
We will now linearize Eq. (\ref{fkk_eq}) for the case of a weak external potential ${\hat U}$ around the field-free solution $f_0$,
\begin{equation}
f_{k,k'}=f^0_{k,k'} + \delta f_{k,k'}, \quad \mbox{with}\quad |\delta f_{k,k}|<< f_k.
\label{lin_f0}
\end{equation}
Without external field it obeys a spatially homogeneous equation, i.e. $f^0_{k,k'}=f_k\delta_{k,k'}$,
\begin{eqnarray}
 i\hbar \frac{\partial f^0_{k}}{\partial t} = I^0_{k}, 
\label{f0_eq}
\end{eqnarray}
where the other terms in Eq. (\ref{fkk_eq}) vanish due to homogeneity.
In the collisionless case, the unperturbed solution is taken as the equilibrium distribution -- in the quantum case, as Fermi distribution,
\begin{equation}
 f_k^{EQ}=\left[ e^{\beta \left( \epsilon_k - \mu\right) } +1 \right]^{-1},
\label{fermi}
\end{equation}
where $\beta= 1/k_BT$ and $\mu(n,T)$ is the chemical potential. This is correct for Markovian collision integrals which vanish for Fermi-Dirac functions, i.e. $I^0_k[f_k^{EQ}]\equiv 0$, but neglect nonideality effects \cite{bonitz-book}. In principle, instead of Eq. (\ref{fermi}) any other stationary solution is possible. For generalized non-Markovian collision integrals one first has to solve for the correlated distribution function, as it can be done e.g. within nonequilibrium Green's functions theory, e.g. \cite{kwong00}.

The next step is the solution of the linearized equation for $\delta f_{k,k}$ which is driven by an external electromagnetic perturbation and, in the collisionles case, gives rise to the Lindhard (RPA) dielectric function (DF) or, in the classical limit, to the Vlasov DF. In case of a monochromatic excitation the perturbation of the distribution oscillates resonantly with the field without irreversible relaxation. There is no back reaction on the function $f_k$.

Here, we want to go beyond this approximation and include the effect of collisions into the dielectric properties and in the induced electrostatic potential. Thereby we will assume a Markovian collision term which allows us to use Eq. (\ref{fermi}) for the unperturbed solution.
We first write down the equation for the perturbation $\delta f_{k,k}$ which is obtained by linearizing Eq. (\ref{lin_f0}) around the homogeneous field-free solution $f_k$,
\begin{eqnarray}
\label{lin_f1}
 i\hbar \frac{\partial \delta f_{k,k'}}{\partial t} - (\epsilon_k - \epsilon_{k'})\delta f_{k,k'} 
-\delta U^{{\rm eff}}_{k,k'}\cdot (f_{k'}-f_k) &=& \delta I_{k,k'}, \\
\lim_{t\to -\infty} f_{k,k'}(t) & = & f_k \delta_{k,k'}.
\nonumber
\end{eqnarray}
Here, $\delta U^{{\rm eff}}=U+\delta W_{ee}$ is obtained by replacing $f_{k,k'}$ by $\delta f_{k,k'}$, whereas $\delta I$ is 
obtained by keeping in all appearances of the electron distribution functions only terms of first order in $\delta f$ \cite{kwong00}. Due to linearity, this equation is conveniently analyzed in Fourier space. Assuming a monochromatic perturbation, $U({\bf r},t)=U_q(\omega)e^{-i(\omega+i\delta) t +iqr}$ the same time and space dependences occure in all terms. Here $\delta$ is an infinitesimal positive real number assuring that the potential vanishes at $t\rightarrow -\infty$. 

To relate the wave vector $q$ of the spatial modulation to the momenta $k, k'$ we introduce center and relative momenta by 
\begin{eqnarray}
 Q &=& \frac{k+k'}{2}, \qquad q = k-k', \quad \mbox{or, vice versa,} \quad \\
 k &=& Q+\frac{q}{2}, \qquad k' = Q-\frac{q}{2}.
\nonumber
\end{eqnarray}
Then, Eq. (\ref{lin_f1}) can be written in Fourier representation as
\begin{eqnarray}
\label{lin_f1_f}
 \left[ \hbar(\omega+i\delta) - (\epsilon_k - \epsilon_{k'})\right] \delta f_{k,k'}  
-\delta U^{{\rm eff}}_{q}(\omega)(f_{k'}-f_k) &=& \delta I_{q}(\omega),
\end{eqnarray}
where the common exponential factor has been cancelled.

\subsection{Collision integral in relaxation time approximation}
If collisions are present they give rise to relaxation to a stationary state
which will be achieved for times exceeding the relaxation time $\tau$ of the system. In the presence of a monochromatic excitation, this asymptotic solution will be modulated by the field and will not 
conoincide with the Fermi function (\ref{fermi}). It can be constructed as a sum of the stationary solutions of Equations (\ref{f0_eq}) -- i.e. the Fermi function -- and the asymptotic solution of Eq. (\ref{lin_f1}). The latter is readily obtained by putting the time derivative to zero and requiring the collision integral to vanish, $\delta I_{k,k'}[\delta f^{\infty}]\equiv 0$. Correspondingly, in the Fourier representation (\ref{lin_f1_f}), we consider the limiting case $\omega \rightarrow 0$
and obtain the stationary asymptotic solution:
\begin{eqnarray}
\label{f1_inf}
 \delta f^{\infty}_{k,k'} = 
\frac{f_{k}-f_{k'}}{\epsilon_k - \epsilon_{k'}}\,\delta U^{{\rm eff},\infty}_{k,k'}, \quad \omega = 0.
\end{eqnarray}
The superscript ``$\infty$'' of the effective potential indicates that the asymptotic solution 
$\delta f^{\infty}$ has to be inserted but also that the static limit of the effective potential has to satisfy an additional consistency condition. As we will see below in Sec. \ref{continuity_ss} this is related to fulfillment of sum rules and conservation laws.

Thus we found the asymptotic solution of the full electron distribution function
\begin{equation}
 \lim_{t\to +\infty} f_{k,k'}(t)  =  f_k \delta_{k,k'} + \delta f^{\infty}_{k,k'} \delta_{\omega,0}
\label{f_inf_0}
\end{equation}
and, for $\omega \ne 0$,
\begin{equation}
 \lim_{t\to +\infty} f_{k,k'}(t)  =  f_k \delta_{k,k'} + \delta f_{k,k'}(\omega)e^{-i\omega t}.
\label{f_inf}
\end{equation}

Since $\delta f$ is a small correction, the stationary result (\ref{f_inf_0}) can, in linear order, be included into the Fermi function giving rise to a local Fermi distribution
$ f^{EQ}({\bf r})=\left[ e^{\beta \left( \epsilon_k - \mu-\delta \mu({\bf r})\right) } +1 \right]^{-1},
$ where $\delta \mu({\bf r}) = \delta U^{{\rm eff},\infty}({\bf r})$ \cite{mermin70}, consistent with thermodynamic stability in an external field, but this is not necessary for the derivations below.

After having found the asymptotic solution $\delta f^{\infty}_{k,k'}$ of the distribution function we can construct the collision integral $\delta I$ which drives the relaxation towards this result.
A simple static model which has this property is the standard relaxation time approximation. We recall the result for the spatially homogeneous case (we drop all arguments),
\begin{equation}
\frac{df}{dt}=I^{RTA}\equiv - \frac{1}{\tau}\left(f - f^{EQ}\right), \quad f(0)=f_0. 
\label{rta}
\end{equation}
The solution of this equation together with the initial condition at $t=0$ is
$f(t)=f_0 e^{-t/\tau} + f^{EQ}[1-e^{-t/\tau}]$, showing the decay of the initial state and the approach to the  asymptotic state. Using this result we can now construct the collision integral $\delta I$ of Eq. (\ref{lin_f1_f}) in RTA. Replacing in Eq. (\ref{rta}) $f$ by $\delta f$ and $f^{EQ}$ by $\delta f^{\infty}_{k,k'}$ [the function $f_k$ cancels] and multiplying by $i\hbar$ we obtain
\begin{eqnarray}
\label{dI_rta}
 \delta I^{RTA}_{k,k'}(\omega) = - \frac{i\hbar}{\tau}\left\{ \delta f_{k,k'}(\omega)-
\frac{f_{k}-f_{k'}}{\epsilon_k - \epsilon_{k'}}\,\delta U^{{\rm eff},\infty}_{k,k'}
\right\}
\end{eqnarray}
Here, $\tau$ is the relaxation time which has to be computed from a separate kinetic theory or taken from experiment. We use a simple static approximation where $\tau$ is frequency independent which is consistent with the asymptotic result (\ref{f1_inf}) which was obtained for $\omega = 0$.
Recall that $\delta U^{{\rm eff},\infty}_{k,k'}$ still has to be determined.  

\subsection{Distribution function $\delta f$ with collisions}\label{continuity_ss}
Using the result (\ref{dI_rta}) in Eq. (\ref{lin_f1_f}) we are now ready to explicitly compute the perturbation of the distribution function. A straightforward calculation gives 
\begin{eqnarray}
\label{df_solution}
 \delta f_{k,k'}(\omega) = \left\{ \delta U^{{\rm eff}}_{k,k'}(\omega) - i\hbar \nu \frac{\delta U^{{\rm eff},\infty}_{k,k'}}{\epsilon_k - \epsilon_{k'}} \right\}
\frac{f_{k'}-f_{k}}{\hbar(\omega+i\nu)-[\epsilon_k - \epsilon_{k'}]} \,,
\end{eqnarray}
where we used $\delta \rightarrow 0$ due to the existence of a finite collisional damping $\nu = \tau^{-1}$. This result is a straightforward extension of the collisionless random phase approximation. 
Scattering effects (terms proportional to $\nu$) are contained in two places: first, the frequency in the denominator is replaced by a complex frequency and, second, there appears an additional contribution proportional to $\delta U^{{\rm eff},\infty}$ in the numerator which renormalizes the Fourier component of the effective potential. We now turn to the computation of this latter term.

To this end we consider the local particle conservation law, i.e. the continuity equation
\begin{eqnarray}
 \frac{\partial n({\bf r},t)}{\partial t} + {\rm div} \,{\bf j}({\bf r},t) = 0,
\label{cont}
\end{eqnarray}
where density and current density of the electrons are related to the distribution functions by 
\begin{eqnarray}
 n({\bf r},t) &=& 2 \int \frac{d^3p}{(2\pi\hbar)^3} f_e({\bf p}, {\bf r},t), \\
 {\bf j}({\bf r},t) &=& 2 \int \frac{d^3p}{(2\pi\hbar)^3} \frac{{\bf p}}{m_e}f_e({\bf p}, {\bf r},t),
\end{eqnarray}
and the prefactor $2$ accounts for the spin summation. Now, in the field-free case, the distribution function is time-independent and spatially homogeneous and does not contribute to the continuity equation. In contrast, in the presence of the external field $U$ there is a time and space dependent contribution $\delta f$ which contributes to $n$ and ${\bf j}$. Thus, the continuity equation becomes, after transformation to Fourier space,
\begin{eqnarray}
 \omega \delta n_q(\omega) = {\bf q}\cdot \delta{\bf j}_q(\omega),
\label{cont_f}
\end{eqnarray}
where 
\begin{eqnarray}
 \delta n_q(\omega) &=& 2 \int \frac{d^3Q}{(2\pi)^3} \,\delta f_{Q+\frac{q}{2}, Q-\frac{q}{2}}(\omega), \\
 \delta{\bf j}_q(\omega) &=& 2 \int \frac{d^3Q}{(2\pi)^3}\frac{\hbar{\bf Q}}{m_e} \,
\delta f_{Q+\frac{q}{2}, Q-\frac{q}{2}}(\omega).
\end{eqnarray}

We will now verify that local particle conservation (\ref{cont_f}) is satisfied by our solution which gives the required condition for $U^{{\rm eff},\infty}$. To compute the density and current from the solution (\ref{df_solution}) requires to perform integrations over $Q$. To shorten the notation we introduce the following relevant integrals ($n=0,1,\dots$)
\begin{eqnarray}
 \Pi_n(q,{\hat \omega}) &=& 2\int \frac{d^3Q}{(2\pi)^3}\left(\frac{\hbar {\bf Q}}{m_e}\right)^{n}
\frac{f_{+}-f_{-}}{\hbar {\hat\omega}-[\epsilon_{+} - \epsilon_{-}]},
\\
 \Pi_{\nu, n}(q,{\hat \omega}) &=& 2\int \frac{d^3Q}{(2\pi)^3}\left(\frac{\hbar {\bf Q}}{m_e}\right)^{n}
\frac{1}{\epsilon_{+} - \epsilon_{-}}\times \frac{f_{+}-f_{-}}{\hbar {\hat\omega}-[\epsilon_{+} - \epsilon_{-}]},
\end{eqnarray}
where we introduced the abbreviation for the complex frequency ${\hat \omega}=\omega+i\nu$. In the collisionless limit, $\nu \rightarrow \delta$, and the $\Pi_n$ become (moments of) the retarded polarization function 
$\Pi^R_n$ \cite{bonitz-book}. Further, we introduced the short notation $Q\pm\frac{q}{2}\rightarrow \pm$. Note that the integrals with odd powers are vectors in the direction of ${\bf Q}$.
We these definitions the Fourier components of the density and current density become (due to homogeneity, $\delta U^{{\rm eff}}_{k,k'}=\delta U^{{\rm eff}}_{q}$)
\begin{eqnarray}
\delta n_q(\omega) &=& \Pi_0(q,{\hat \omega}) \delta U^{{\rm eff}}_{q}(\omega) 
-i\hbar \nu \,\delta U^{{\rm eff}, \infty}_{q}\Pi_{\nu,0}(q,{\hat \omega}),
\label{dn1}
\\
\delta{\bf j}_q(\omega) &=& \Pi_1(q,{\hat \omega}) \delta U^{{\rm eff}}_{q}(\omega) 
-i\hbar \nu \,\delta U^{{\rm eff}, \infty}_{q}\Pi_{\nu,1}(q,{\hat \omega}).
\label{dj1}
\end{eqnarray}
We now transform $\hbar {\bf q}$ times the integral $\Pi_1$, by adding and subtracting under the integral $\hbar {\hat \omega}$. Taking into account that $\epsilon_+-\epsilon_- = \hbar^2 {\bf Q}\cdot {\bf q}/m_e$, we obtain the identity
\begin{equation}
 \hbar {\bf q}\Pi_1(q,{\hat \omega}) = - 2\int \frac{d^3Q}{(2\pi)^3} (f_- - f_+)
+(\hbar \omega + i\hbar \nu)\Pi_0(q,{\hat \omega}).
\label{p1_identity}
\end{equation}
Assuming that the field-free distribution depends only on the modulus of the momentum, i.e.
$f_{-{\bf k}}=f_{{\bf k}}$ the integrals over $f_-$ and $f_+$ cancel. The same transformation is possible for the integral $\Pi_{\nu,1}$ with the result
\begin{equation}
 \hbar {\bf q}\Pi_{\nu,1}(q,{\hat \omega}) = \Pi_0(q,0)
+(\hbar \omega + i\hbar \nu)\Pi_{\nu,0}(q,{\hat \omega}).
\label{pnu1_identity}
\end{equation}
Collecting the results (\ref{p1_identity}) and (\ref{pnu1_identity}) together we may rewrite the expression for the current density, Eq. (\ref{dj1})
\begin{eqnarray}
\hbar{\bf q}\delta{\bf j}_q(\omega) &=& 
\hbar(\omega+i\nu)\Pi_0(q,{\hat \omega})\delta U^{{\rm eff}}_{q}(\omega) 
\nonumber\\
&& -i\hbar \nu \,\left[\Pi_0(q,0)+(\hbar \omega + i\hbar \nu)\Pi_{\nu,0}(q,{\hat \omega}) \right]\delta U^{{\rm eff}, \infty}_{q} = 
\nonumber\\
&=& \hbar \omega \delta n_q(\omega) + 
\nonumber\\
&&+ i\hbar \nu \left\{\Pi_0(q,{\hat \omega})\delta U^{{\rm eff}}_{q}(\omega) -
U^{{\rm eff}, \infty}_{q}\left[ \Pi_0(q,0)+ i\hbar \nu\Pi_{\nu,0}(q,{\hat \omega}) \right]\right\}.
\label{dj2}
\end{eqnarray}
Evidently, the continuity equation (\ref{cont_f}) is fulfilled if the terms on the last line (in the curley brackets) vanish. From this we find the condition for the asymptotic value of the effective potential
\begin{equation}
 U^{{\rm eff}, \infty}_{q}=\frac{\Pi_0(q,{\hat \omega})\delta U^{{\rm eff}}_{q}(\omega)}{\Pi_0(q,0)+ i\hbar \nu \Pi_{\nu,0}(q,{\hat \omega})} = \frac{\delta n_q(\omega)}{\Pi_0(q,0)}.
\label{u_inf}
\end{equation}
To obtain the last equality we used $\Pi_0(q,{\hat \omega})\delta U^{{\rm eff}}_{q}(\omega)=\delta n_q(\omega)+i\hbar\nu \Pi_{\nu,0}(q,{\hat \omega})U^{{\rm eff}, \infty}_{q}$. With this the problem has been solved and the perturbation of the electron distribution function (\ref{df_solution}) has been obtained explicitly:
\begin{eqnarray}
\label{df_solution_final}
 \delta f_{k,k'}(\omega) = \left\{ \delta U^{{\rm eff}}_{q}(\omega) - i\hbar \nu \frac{\delta n_q(\omega)}{\Pi_0(q,0)[\epsilon_k - \epsilon_{k'}]} \right\}
\frac{f_{k'}-f_{k}}{\hbar(\omega+i\nu)-[\epsilon_k - \epsilon_{k'}]}\: \,.
\end{eqnarray}

\section{Quantum dielectric function containing collisions}
To compute the dielectric function we first have to obtain an explicit result for the density fluctuation $\delta n$. Using (\ref{dn1}) and (\ref{u_inf}) or integrating (\ref{df_solution_final}), we can write
\begin{eqnarray}
\delta n_q(\omega) &=& \Pi_0(q,{\hat \omega}) \delta U^{{\rm eff}}_{q}(\omega) 
-i\hbar \nu \,\frac{\Pi_{\nu,0}(q,{\hat \omega})}{\Pi_0(q,0)}\delta n_q(\omega).
\label{dn2}
\end{eqnarray}
Recalling the definition of the effective potential, 
$\delta U^{{\rm eff}}_{q}(\omega)=\delta U_q(\omega)+V_q\delta n_q(\omega)$, 
and introducing the short notation ${\tilde\Pi}_{\nu,0}(q,{\hat \omega}) \equiv \Pi_{\nu,0}(q,{\hat \omega})/\Pi_0(q,0)$ we can solve for $\delta n_q$:
\begin{eqnarray}
\delta n_q(\omega) = \frac{\Pi_0(q,{\hat \omega}) }{1-V_q\Pi_0(q,{\hat \omega})+i\hbar\nu {\tilde \Pi}_{\nu,0}(q,{\hat \omega})}\:\delta U_{q}(\omega).
\label{dn3}
\end{eqnarray}
The inverse dielectric function is defined as \cite{bonitz-book}
\begin{equation}
 \epsilon^{-1}_q(\omega)=\frac{\delta \delta U^{{\rm eff}}_{q}(\omega)}{\delta U_q(\omega)}
\label{df}
\end{equation}
and is obtained by inserting the solution (\ref{dn3}) into the effective potential with the result 
\begin{equation}
\delta U^{{\rm eff}}_{q}(\omega) = 
\frac{1+i\hbar \nu {\tilde \Pi}_{\nu,0}(q,{\hat \omega})}
{1-V_q\Pi_0(q,{\hat \omega}) + i\hbar\nu {\tilde \Pi}_{\nu,0}(q,{\hat \omega})}\,U_q(\omega). 
\label{ueff_result} 
\end{equation}
Performing the derivative with respect to $\delta U$ and inverting the result, we obtain
the dielectric function
\begin{equation}
\epsilon_q(\omega)=
1 - \frac{V_q\Pi_0(q,{\hat \omega})}
{1+i\hbar \nu {\tilde \Pi}_{\nu,0}(q,{\hat \omega})}
\label{df_result} 
\end{equation}
An alternative way of writing this result is to eliminate the function $\Pi_{\nu,0}$ by using the identity $\hbar \omega\Pi_{\nu,0}(q,{\hat \omega})=\Pi_0(q,{\hat \omega})-\Pi_0(q,0)$. In this case, in the expression for the dielectric function one can make the replacement 
${\tilde \Pi}_{\nu,0}(q,{\hat \omega})=[\Pi_{0}(q,{\hat \omega})/\Pi_0(q,0)-1]/\hbar\omega$.

This way we have succeeded to derive a dielectric function which contains collision effects in a relaxation time approximation with a static collision frequency $\nu$. This is the result of Mermin \cite{mermin70} which had been derived before for classical plasmas by Rostoker and Rosenbluth, e.g. \cite{alexandrov}. 

\subsection{Further improved dielectric functions}
Further improvements of the Mermin result have been considered by various groups. R\"opke et al. have derived a Mermin-type expression which, besides particle conservation contains energy conservation \cite{selchow02}. However, they found that the effect was small. Another modification by this group was to include a frequency dependent collision frequency into the relaxation time collision integral \cite{millat03}. Finally, we mention that a selfconsistent nonequilibrium calculation within Nonequilibrrium Green's functions which fully included sum rule preservation has been recently performed \cite{kwong00}.

\subsection{Dynamical screening and wake effects}
The main motivation to include dynmical screening of the interaction between heavy particles in a two-component plasma is its importance for nonequilbrium situations. One such case is the existing of streaming light particles which causes wake effects which have a dramatic effect on the arrangement of heavy particles, e.g. dust particles in a complex plasma. This was discussed in detail by Joyce and Lampe, cf. \cite{joyce02} and references therein. Wake effects in a quantum plasma have also been considered by one group \cite{abril98} who found an important influence on stopping power of ions in a polarizable medium. However, these were only single particle effects.

\section{Dynamically screened ion-ion pair potential}
Let us now compute the potential of a moving charged particle taking into account the dielectric properties of the plasma. We start with the case of a classical charged particle, e.g. \cite{alexandrov}, and then generalize the result to quantum particles.

\subsection{Potential of a moving classical particle}
The Poisson equation for a polarizable medium reads 
\begin{equation}
 {\rm div} {\bf D}({\bf r},t) = e_a n_a({\bf r},t),
\label{d}
\end{equation}
which in Fourier space reads $i{\bf k}\cdot{\bf D}_k(t)=e_a n_k(t)$. The electrostatic potential $\phi$ created by the charge density on the right is ${\bf E}({\bf r})=-\nabla \phi({\bf r})$, corresponding, in Fourier space to ${\bf E}_k=-i{\bf k}\cdot \phi_k$. Together with the electrodynamic definition of the dielectric tensor, $D_{k,i}=\sum_j\epsilon_{k,ij} E_{k,j}, \quad i,j=1,2,3$ and Eq. (\ref{d}) we obtain
\begin{equation}
 \phi_k(\omega) = \frac{e_a n_{ak}(\omega)}{\sum_{ij} k_ik_j \epsilon_{k,ij}(\omega)}.
\label{phi_k}
\end{equation}
Consider now the case of a classical point charge ``a'' with initial position ${\bf r}(0)={\bf r}_{0a}$, moving with constant velocity ${\bf v}_a$ (relative to the carriers creating the dielectric function). Then
\begin{equation}
n_a({\bf r},t) = \delta[{\bf r}-{\bf r}_{0a} - {\bf v}_at],
\label{n_point}
\end{equation}
with the Fourier representation
\begin{equation}
n_{ak}(\omega) = 2\pi e^{i{\bf k}{\bf r}_{0}}\delta[\omega - {\bf k}{\bf v}_a].
\label{nk_point}
\end{equation}
Inserting this result into (\ref{phi_k}) and performing the back transform we obtain
\begin{equation}
 \phi({\bf r}-{\bf r}_{0a};{\bf v}_a) = e_a\int \frac{d^3 k}{(2\pi)^2}\frac{e^{i{\bf k}({\bf r}_{0a}-{\bf r})}}{\sum_{ij} k_ik_j \epsilon_{k,ij}({\bf k}{\bf v}_a)}.
\label{phi_r}
\end{equation}
Thus, the potential of a classical point particle moving with a constant velocity is time-independent, only a single frequency component, $\omega={\bf k}{\bf v}_a$, is present in the spectrum.
For the special case of an isotropic medium, $\epsilon_{ij}$ has only two independent components. 
Concentrating on longitudinal plasma oscillations we can replace $\epsilon_{ij}\rightarrow k_ik_j\epsilon/k^2$, and the potential (\ref{phi_r}) becomes
\begin{equation}
 \phi({\bf r}-{\bf r}_{0a};{\bf v}_a) = \int \frac{d^3 k}{(2\pi)^2}\frac{e_a}{k^2}
\frac{e^{i{\bf k}({\bf r}_{0a}-{\bf r})}}{\epsilon_{k}({\bf k}{\bf v}_a)}.
\label{phi_rl}
\end{equation}

This result may be immediately generalized to the case of many particles. Indeed, due to linearity of Maxwell's equations, the resulting total potential is simply the sum of all potentials of the type (\ref{phi_r}), i.e. $\phi^{tot}({\bf r})=\sum_{a=1}^N \phi({\bf r}-{\bf r}_{0a};{\bf v}_a)$.

\subsection{Potential of moving quantum particles}
In case the charge density is created by quantum particles which are not point like or by an ensemble of many particles the particle density is expressed, e.g. via the single-particle distribution function, 
\begin{equation}
n({\bf r},t)=2 \int\frac{d^3p}{(2\pi\hbar)^3}f({\bf p},{\bf r},t).
\label{n_f}
 \end{equation}
On the other hand we can consider the result (\ref{phi_r}) in the continuum limit of a superposition of infinitesimal charges. Then the contribution $d\phi$ from the charge $dq({\bf r})$ inside a small volume $dV$ around point ${\bf r}$ is proportional to $dq({\bf r})/dV=en({\bf r})$, and the total electrostatic potential is 
\begin{equation}
 \phi({\bf r};{\bf v}_a) = \int d^3r'\int \frac{d^3 k}{(2\pi)^2}\frac{e n({\bf r}')}{k^2}
\frac{e^{i{\bf k}({\bf r}'-{\bf r})}}{\epsilon_{k}({\bf k}{\bf v}_a)},
\label{phi_cont}
\end{equation}
where $n$ has to be computed from the distribution function according to (\ref{n_f}), and it has been assumed that all particles stream with the same velocity ${\bf v}_a$. This result is applicable to classical and quantum systems equally and provides, in particular, the induced potential of an ensemble of degenerate electrons with $e$ characterized by a distribution function $f_e$. 

\subsection{Ion-ion pair potential}

We can now return to the ion dynamics studied in section \ref{sec:eqsl}. There the effect of the electrons was comprised in the electrostatic potential $\phi_e$, cf. Eq. (\ref{fi_eqcl}). 
If the electron density is spatially homogeneous, the particles will produce a homogeneous potential $\phi({\bf r})=const$, as can be seen from Eq. (\ref{phi_cont}), and will not exert a force on this ions.
In an external potential the electron density maybe spatially modulated but this will be a small effect. Therefore, the main effect of the electrons is polarization of the plasma medium via the dielectric function (\ref{df_result}). In this case, the electrons will renormalize (screen) any other potential created by charged particles, in particular the ion-ion interaction, term $V_i^Q$ in Eq. (\ref{fi_eqcl}) will be screened, i.e. $V_i^Q \rightarrow {\tilde V}_i^Q$. Due to the linear approximation, we may write
\begin{equation}
 {\tilde V}_i^Q({\bf r})=\sum_{1\le k<l\le N_i} {\tilde V}^Q_{ii}({\bf r}_k-{\bf r}_l).
\end{equation}
The corresponding force term in the Newton's equation for the $k$-th ion will be minus the gradient of
\begin{equation}
 e_k\phi_{tot}({\bf r}_k)=\sum_{1\le 1\le N_i, l\ne k} {\tilde V}^Q_{ii}({\bf r}_k-{\bf r}_l)
=e_k\sum_{1\le 1\le N_i, l\ne k}\phi_l({\bf r}_k-{\bf r}_l).
\end{equation}
The screened potential created by the $l$-th ion which moves with velocity ${\bf v}_l(t)$ (relative to the electrons) then follows directly from our result (\ref{phi_rl}). Note the direction dependence of the potential and the dynamically screened pair potential.
If the electrons stream with a constant velocity ${\bf u}_e$ this potential becomes 
\begin{equation}
 \phi_l({\bf r}-{\bf r}_{0l};t) = e_l\int \frac{d^3 k}{(2\pi)^2}\frac{1}{k^2}
\frac{e^{i{\bf k}({\bf r}_{0l}-{\bf r})}}{\epsilon_{k}\left({\bf k}\cdot\left[{\bf v}_l(t)-{\bf u}_e\right]\right)}
\label{screened_ion_pot}
\end{equation}
and may be time-dependent through the time dependence of the velocity $v_l$. Further it is straightforward to include ionization and recombination effects through a time dependence of the charges $e_k$ and $e_l$. 

\subsection{Limiting cases}\label{limits_ss}

The classical limit of the potential (\ref{screened_ion_pot}) gives the Rostoker-Rosenbluth result \cite{rostoker} which was used in classical MD simulations of dusty plasmas \cite{joyce02}. The present potential is a straightforward generalization to dynamical screening by quantum particles. The relevance of dynamical screening depends on the particle velocities. If the electrons are at rest, we will have, on average ${\bf u}_e=0$. If further, the ions are at rest or moving very slowly, the frequency argument of the dielectric function vanishes and $\epsilon_k \rightarrow k^2/(k^2+\kappa^2)$. In thermodynamic equilibrium the screening parameter is given by the Thomas-Fermi length, $\kappa^{-1}\approx r_{TF}$. Then the potential (\ref{screened_ion_pot}) contains just the Fourier transform of $(k^2+\kappa^2)^{-1}$ which yields the familiar Yukawa potential. Thus the resulting ion-ion pair potential reduces to the isotropic static Yukawa pair potential $V_{ii}(r)\rightarrow e^2e^{-\kappa r}/r$. Thus the validity of the Yukawa approximation is limited to electrons at rest and to the neglegibility of ion thermal motion as well as of quantum degeneracy effects.

\subsection{Quantum effects in the ion-ion pair potential}\label{q-ion-pot_ss}
Let us briefly comment on the quantum effects in the pair interaction (denoted by the superscript $``Q''$). In using the expression (\ref{phi_rl}) we assumed pointlike ions and neglected any finite ion extension. This can be corrected in the final expression by replacing the Fourier transform of the Coulomb potential, i.e. the factor $1/k^2$ in Eq. (\ref{screened_ion_pot}), by the Fourier transform of the improved Kelbg potential \cite{afilinov_pre04} or of any other appropriate quantum potential.

\section{Discussion}
A new model for the simulation of dense quantum plasmas including dynamical screening of the electrons, partial ionization and strong ion correlations has been developed. It is particularly important for high density low-temperature plasmas, in situations where the ions form liquid or solid-like structures and when the electrons are in nonequilibrium. Typical situations are streaming electrons or plasma instabilities due to fast electrons or electromagnetic fields. The proposed simulation scheme is based on classical molecular dynamics simulations where the dynamical screening effects are incorporated within a linear response approach into the screening of the ion-ion pair interactions. For the screening an extension beyond the (collisionless) RPA model has been used which is due to Mermin and a strict derivation within quantum kinetic theory has been given putting the original result of Ref. \cite{mermin70} on solid ground and critically assessing its scope of applicability.

\end{document}